\documentclass[%
 reprint,
 amsmath,amssymb,
 aps,
]{revtex4-1}

\usepackage{graphicx}
\usepackage{dcolumn}
\usepackage{bm}
\usepackage{epstopdf}
\usepackage{hyperref}
\usepackage[T1]{fontenc}

\begin{document}

\preprint{APS/123-QED}

\title{A Thermodynamic Discriminator for Carbon Nanomaterials}%

\author{Tamoghna Bhattacharya}

 \author{Anjan Kr. Dasgupta}%
 \affiliation{%
 	Department of Biochemistry  \\ University of Calcutta \\ Kolkata 700019 India}
 \altaffiliation[Also at ]{Centre of Excellence in Systems Biology and Biomedical Enineering, University of Calcutta}
 \email{adbioc@caluniv.ac.in}


\begin{abstract}
Interaction between carbon nanomaterials and micellar substrates is studied. A notable observation is the  dependence of nano-surface topology on  thermodynamic  signatures of the carbon nanomaterials e.g., single wall carbon nanotube (SWNT), multiwall carbon nanotube (MWNT) and graphene.  The disruption of the self assembly process while the micelles were converted to monomer has a  unique character in presence of  graphene. This unique behavior follows irrespective of  whether the micelle forming monomer is anionic (Sodium dodecyl sulfate) or cationic(Cetrimonium bromide). The direct measurement of temperature(T) also indicates that T falls monotonically as the micelles are formed in presence of graphene,  this being different in all other cases (SWNT and MWNT). The photon correlation studies indicated formation of smaller and well distributed micelles in contact with graphene ,this being not the case with SWNT and MWNT.  Importantly the free energy change corresponding to the micelle formation has same order of magnitude (-26 to -25 KJ/Mole), the enthalpy showing a nanosurface specific value that varies between -9 to +7 KJ /mole depending on the nature of the nanomaterial and that of the self assembling micellar monomer. The constancy of the free energy and surface dependent vaiations  of enthalpy implies that an entropy enthalpy compensation (free energy being a linear combination of the two) is inevitable in the self assembly process. The micellar cooling induced by graphene further implies a possible potential of the nano-embedded self assembly in fields like energy harnessing and bioenergetic manipulations.
. 
\begin{description}

	\item[PACS numbers] {61.48.De,82.60.Cx,82.70.Uv,82.33.Nq}
    \item[Keywords]{\textbf{} Carbon nanomaterials, enthalpy, coupling, surface 
	topology.}
\end{description}
\end{abstract}

\maketitle


\section{\label{sec:level1}Introduction\protect }
 A  recent article highlights the importance of the topology and geometry in modulating thermodynamic and electrical processes in bulk materials  \cite{gupta2014topological}. Our own work on how topological chirality of nanomaterials can serve as discriminator for  geometrical chirality \cite{bhattacharyya2014chirality} further commends this 'topological twist' of bulk properties like ligand binding. In this paper we explore such surface dependence of  assembly of micellar substances in presence of  multiwall carbon nanotubes (MWCNT)  single wall carbon nanotubes (SWNT) and graphene. It is shown that the surface topology of the nanomaterial has an intricate but specific impact on the thermodynamics of the self assembly process of micellar substances. The packing of micelles \cite{nagarajan2002molecular} is one parameter that is dependent on hydrophobic tail length. The equation expressing the packing parameter is given by \cite{butt2006physics},
 \begin{equation}\label{eq:packing}
 N_s=\frac{V_c}{a\cdot L_c}
 \end{equation}
 where, $V_c,L_c  \& a $ in Eq.(~\ref{eq:packing})respectively represents the volume of the hydrophobic core, area of the cluster and chain length of the tail. In presence of super-hydrophbic  nanomaterials the parameter $N_s$ is likely to change as the each of the three components of the free energy of micelle formation \cite{stokes1997fundamentals},
 \begin{equation}\label{eq:gibbs}
 \Delta G = \Delta G_{HP} +\Delta G_{EL} +\Delta G_{IF} 
 \end{equation}
 
 the subscripts HP, EL and IF in Eq.(~\ref{eq:gibbs}) representing hydrophobic packing interaction in the tail, electrical repulsion between polar heads , and interfacial energy respectively. In presence of super-hydrophobic carbon nanomaterials the interaction between the nano-surface and hydrophobic tail of the miceller object \cite{bhattacharyya2012molecular} are likely to affect the first component. The polar contribution and the interfacial component may also be indirectly affected in presence of a hydrophobic surface. 
 A vast literature has developed on dependence of critical micelle concentration (CMC) and Gibbs free energy of micellization 
of ionic and nonionic surfactant systems \cite{shinoda1954effect,abuin1989study,burrows1995thermodynamics,ballerat1997thermodynamics,freyer2008isothermal,wettig2013thermodynamic,lair2004thermodynamic}. The thermodynamic  parameters (such as enthalpy, entropy etc) and binding behaviors of surfactant micelle have been studied using isothermal titration calorimetry  (ITC)\cite{mazer1982calorimetric,freire1990isothermal}. While the existing literature highlights the  role of surfactants in  dispersing graphene and carbon nanotubes \cite{smith2010importance,moore2003individually,zhao2013adsorption} and in non-covalent functionalization of graphene \cite{rao2010some} there is  a gap in literature concerning  how  carbon nanomaterials modulate the self assembly process of the micellar substrate.  
In some of the reports there has been a confusion over the 'entropy' or 'enthalpy' driven nature of the self assembly process. While some reports such as \cite{moroi1975interrelationship} suggests that there will be always a positive enthalpy change , to make the micelle formation an entropy driven process, there are contradicting reports \cite{qian1996entropy} that micelle formation can have a large negative contribution to the overall free energy change. In this report we find that the sign of enthalpy change is actually a surface toplogy driven process and can change its sign from negative to positive depending on the surface topology of the nanomaterial on whose presence a self assembly process is allowed to occur. \\

\section{\label{sec:level1}Results\protect }
\subsection{\label{sec:level2}Direct Temperature Measurement }

\begin{figure}[htbp]
	\centerline{\includegraphics[width=5in,height=2.17in]{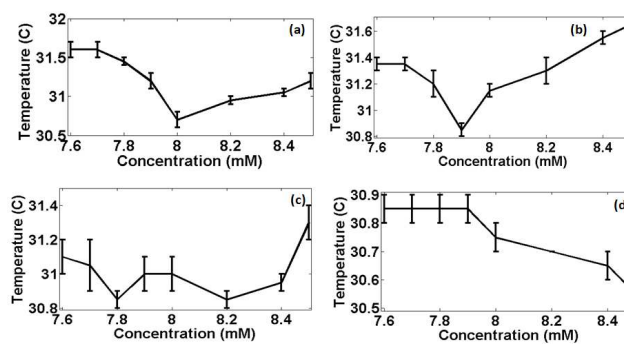}}

	\caption{Variation in temperature  of (a) SDS micelle with water and in presence of (b) SWNTs (c) MWNTs and (d) graphene}
		\label{fig:thermosds}
\end{figure}

\begin{figure}[htbp]	
	\centerline{\includegraphics[width=4in,height=2in]{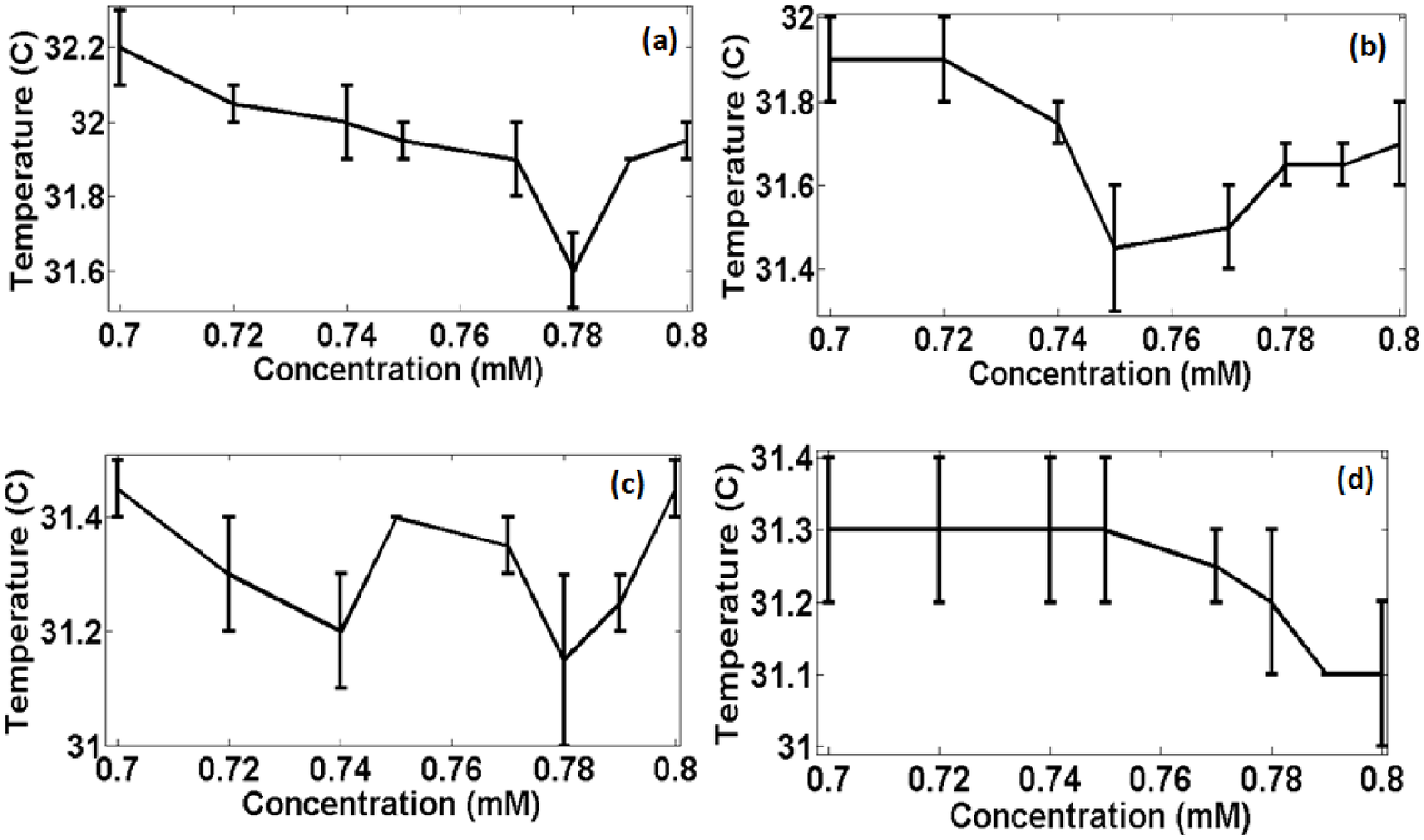}}

	\caption{Variation in temperature  of (a) CTAB micelle with water and in presence of (b) SWNTs (c) MWNTs and (d) graphene}
	\label{fig:thermonano}
\end{figure}

Direct temperature measurement as described in Fig. ~\ref{fig:thermosds} $\&$ Fig. ~\ref{fig:thermonano} reveal a robust thermal character for the self assembly process and their modulation by nanosurface. The micelle formation is always associated with a dip in temperature that is followed by a rise of the same. The exception happens when grapahene is present. Both in Figs. ~\ref{fig:thermosds} $\&$  ~\ref{fig:thermonano} it is found that beyond a break point that is close to the reported CMC values (0.8mM for SDS and 0.78mM for CTAB) the cooling effect is not reversed. In other words the micelles once formed and assumes a certain size are arrested , perhaps on the planar surface of graphene so that no further temperature change occurs. For other cases it seems the micelle formation is followed by formation of super-assemblies leading to further rise of temperature. 
\subsection{\label{sec:level2}Photon Correlation Studies}
\begin{figure}[htbp]
	\centerline{\includegraphics[width=5in,height=2.17in]{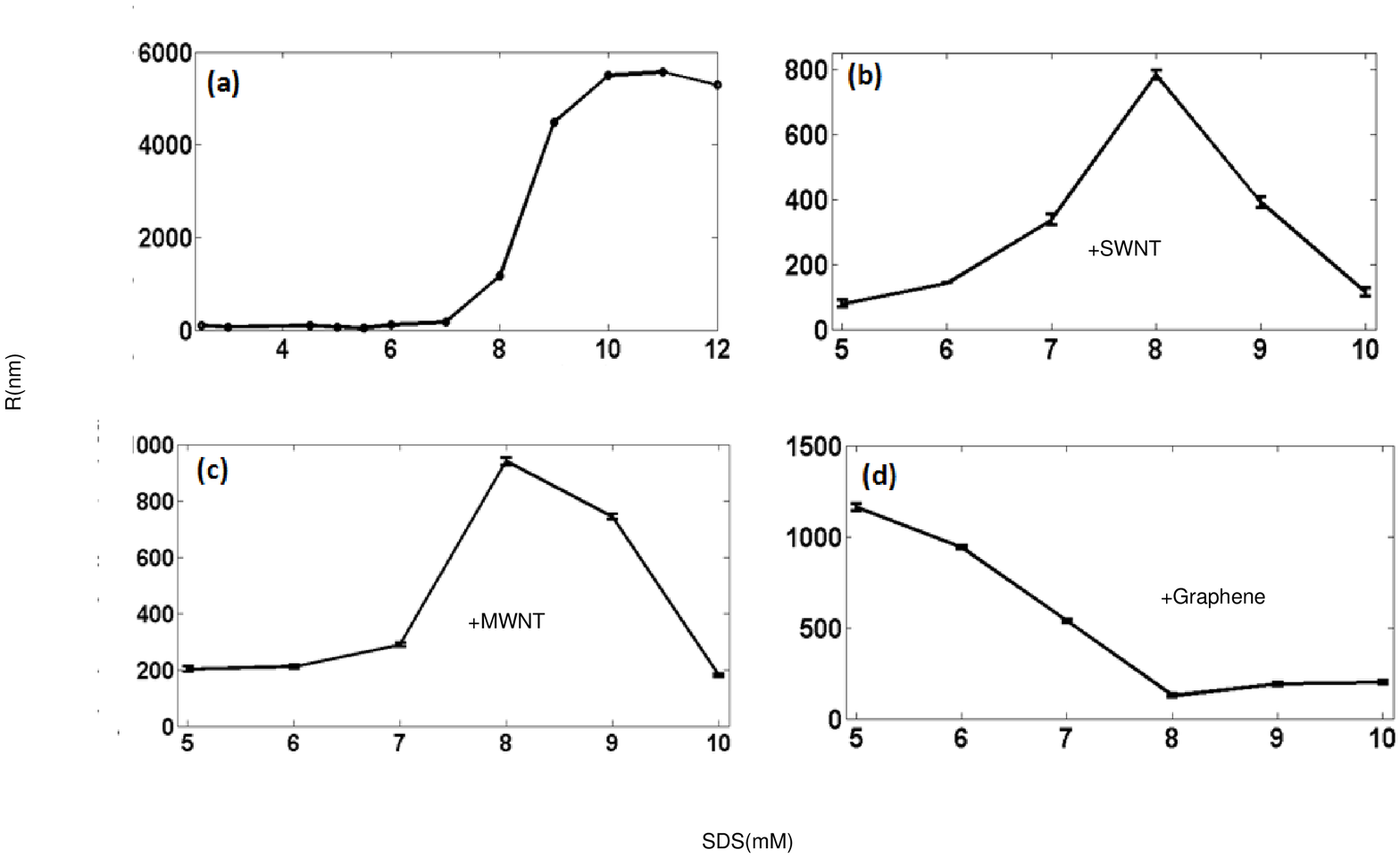}}
	\caption{Variation in hydrodynamic diameter of (a) SDS micelle with water and in presence of (b) SWNTs (c) MWNTs and (d) graphene}
	\label{fig:pcsSDS}
\end{figure}

The thermal measurements tallies well with the hydrodynamic diameter measured  using photon correlation spectroscopy (see Fig. ~\ref{fig:pcsSDS}). There is  an abrupt rise in the diameter value near CMC, that reduces to smaller sizes at higher concentration. Again the exception  comes in presence of graphene , the self assembly process assumes a curious route , the higher concentration of the monomers showing smaller hydrodynamic value. It is possible that the high radii at lower concentration reflects a combined radii of monomers and graphene , the former forming a loose aggregated structure with the  latter. As monomer concentration is increased the monomer monomer interaction becomes more prominant and a stable small miceller structure is phase separated on the graphene surface. The comparison of the SWNT  MWNT and graphene data in Fig. ~\ref{fig:pcsSDS} is also is intriguing as the hydrodynamic diameter plotted against the concentration of monomer carries signature of the surface topology of the respective nanomaterial. The broader (and perhaps binary ) peak(s) of the MWNT may be a reflection of its complex multilayered topology. In contrast the presence of graphene makes the hydrodynamic profile smooth, the  micelle formation in thi scase being on a superhydrophobic planar surface. Similar results are found in case of CTAB (data not shown). 
\subsection{\label{sec:level2}Isothermal Titration Calorimetry (ITC)}
\begin{figure}[htbp]
	\centerline{\includegraphics[width=3.63 in,height=2.79in]{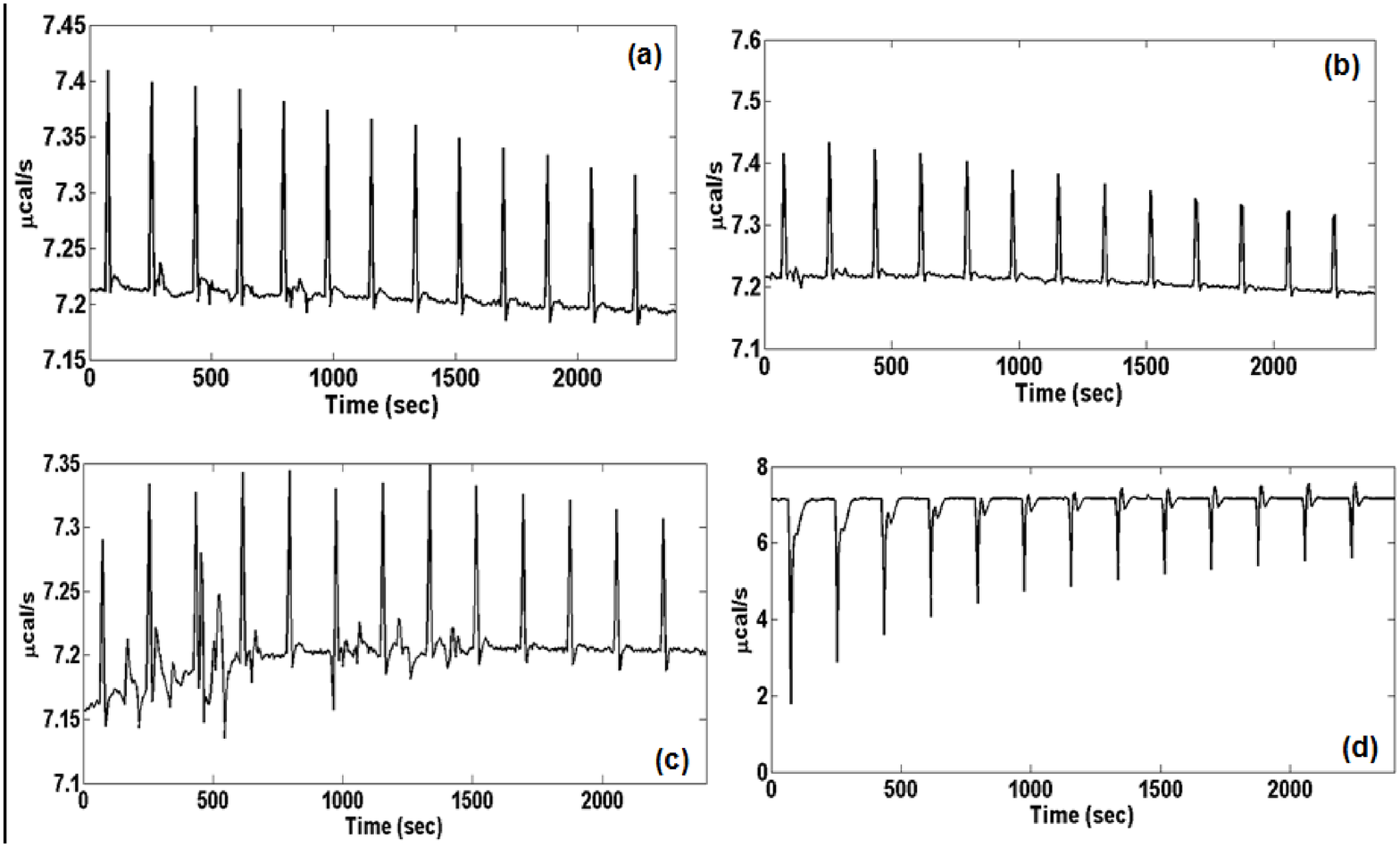}}

	\caption{Change of molar enthalpy of CTAB micelle treated with (a) water and water solubilized (b) SWNTs, (c) MWNTs, (d) graphene}
		\label{fig:ctabitc}
\end{figure}

\begin{figure}[htbp]
	\centerline{\includegraphics[width=3.63 in,height=2.79in]{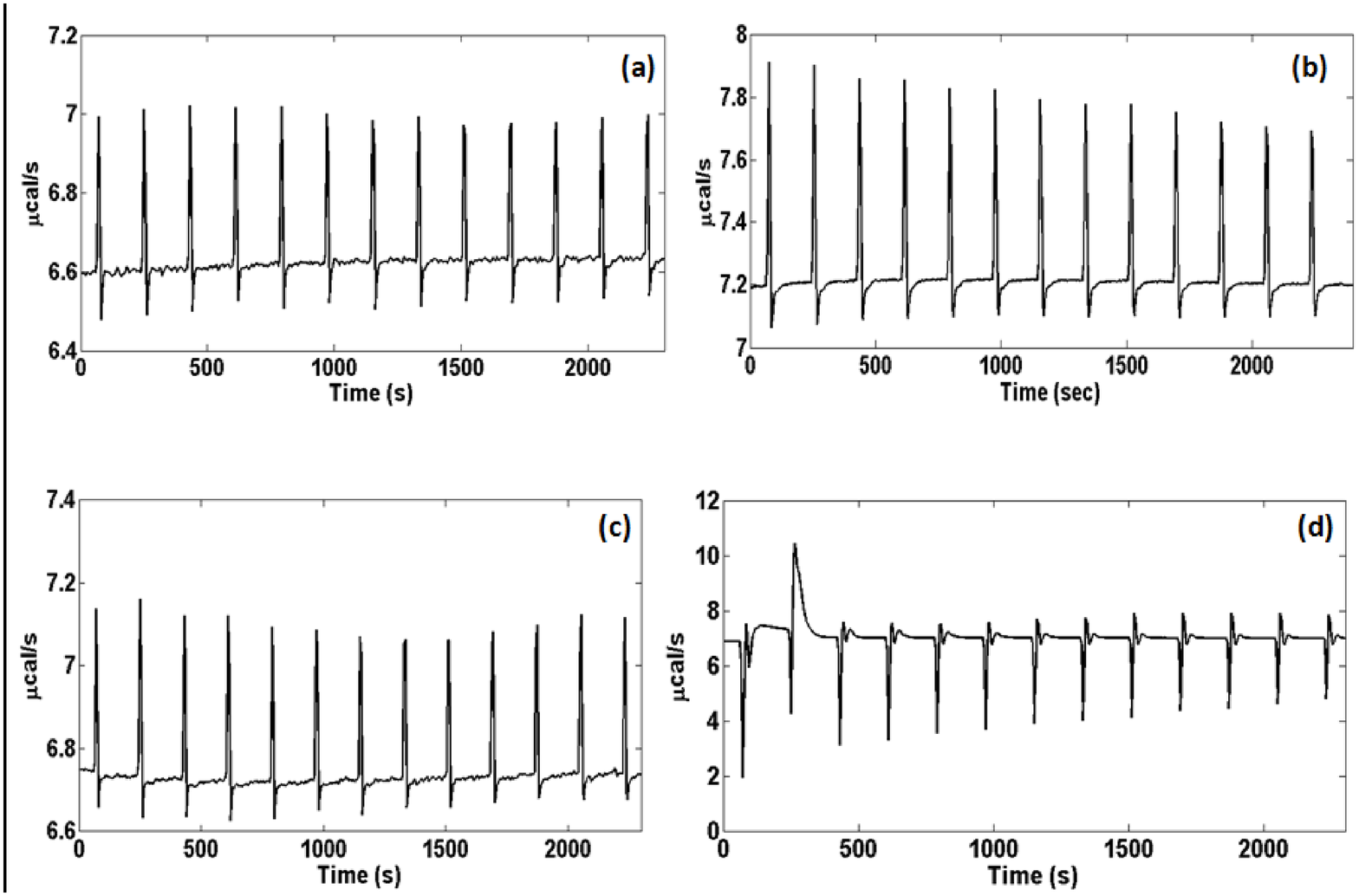}}

	\caption{Change of molar enthalpy of CTAB micelle treated with (a) water and water solubilized (b) SWNTs, (c) MWNTs, (d) graphene}
		\label{fig:sdsitc}
	
\end{figure}

The Fig.s ~\ref{fig:ctabitc}  $\&$ ~\ref{fig:ctab} summarize the isothermal titration characteristics of micellar break down process , in which a micellar solution of CTAB is gradually diluted with water.  SDS (see  Figs. ~\ref{fig:sdsitc} $\&$ Fig. ~\ref{fig:sds} ) also reveal similar ITC profile.  
details of the ITC measurement. The enthalpy value of the miceller substrates in absence of any nanomaterial ,matches with the ones reported in \cite{meagher1998enthalpy} 
\subsection{Entropy Enthalpy Compensation in nanosurface micelle interaction}
\begin{figure}[htbp]
	\centerline{\includegraphics[width=3.46in,height=2.21in]{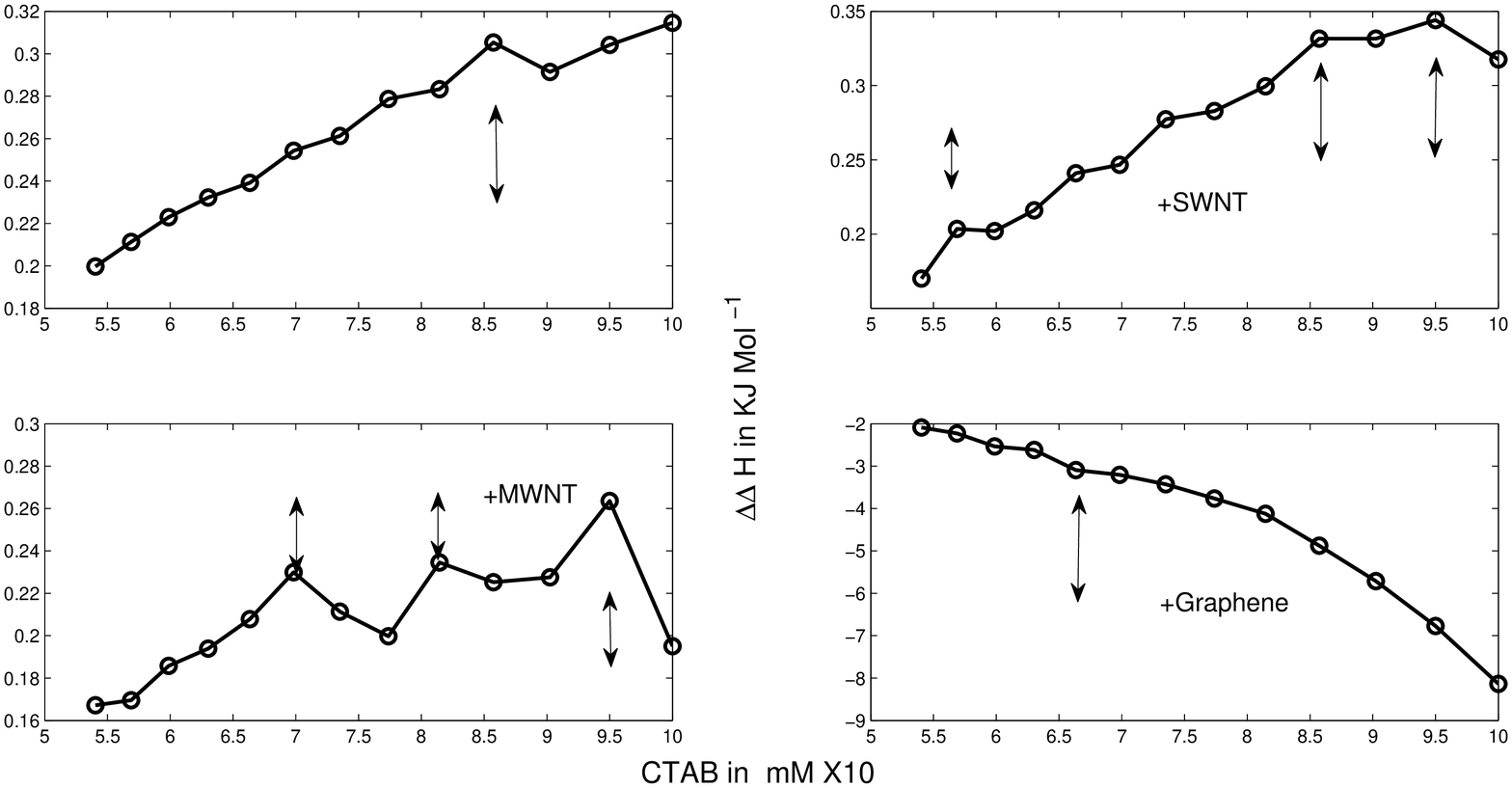}}

	\caption{Enthalpy of micelle breakdown for (a) SDS (b) +SWNT (c) +MWNT (d)+Graphene}
		\label{fig:ctab}
\end{figure}

\begin{figure}[htbp]
	\centerline{\includegraphics[width=3.46in,height=2.21in]{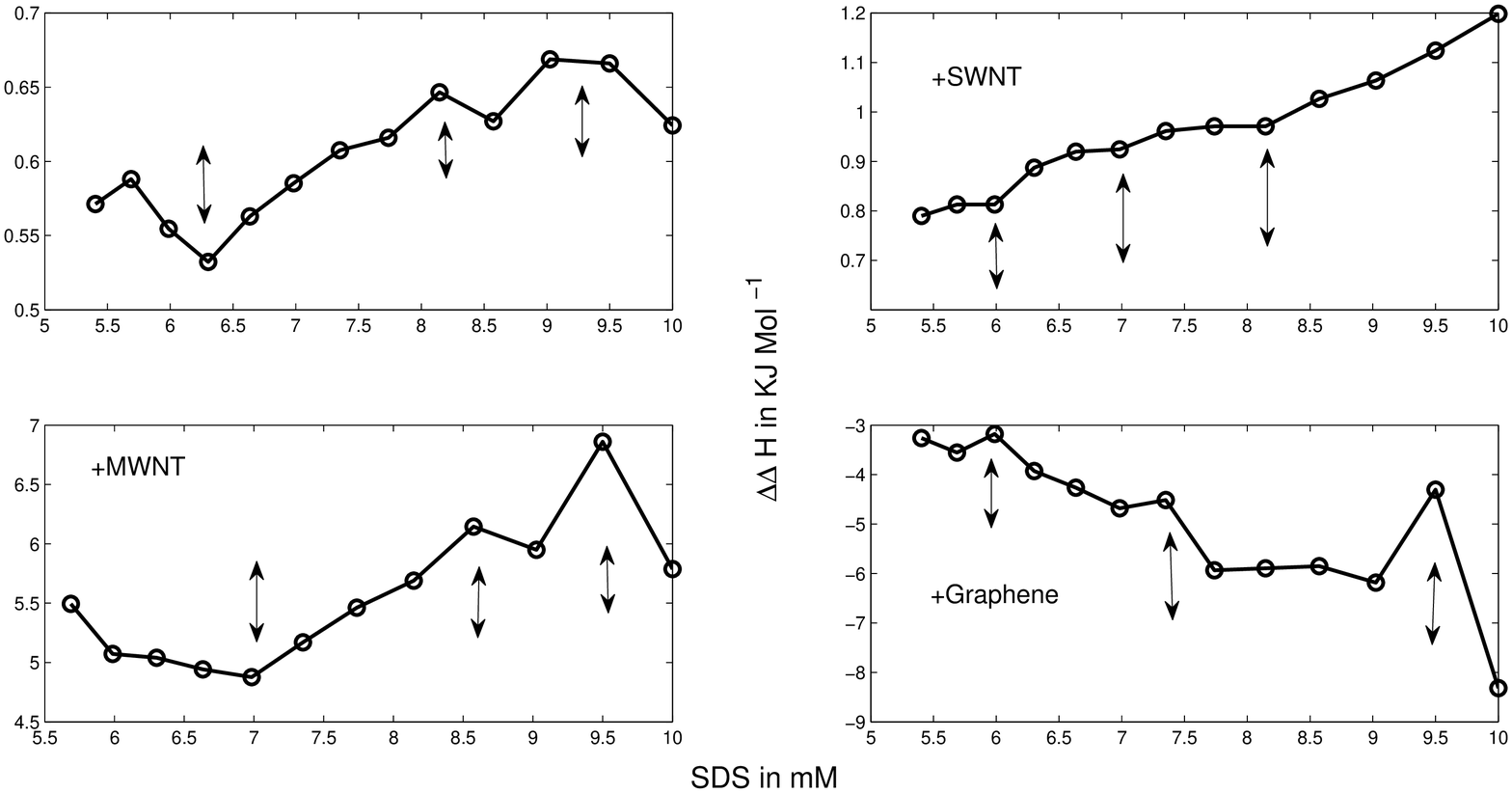}}

	\caption{Enthalpy of micelle breakdown for (a) SDS (b) +SWNT (c) +MWNT (d)+Graphene}
		\label{fig:sds}
\end{figure}

It may be noted that the enthalpic studies reveal a number of break points in each case (CTAB and SDS), the number of break points varying depending on the nanosurface (SWNT, MNTand graphene). If we consider the approximate relation that ,
\begin{equation}
\Delta G = RT ln [CMC]
\label{eq:cmc}
\end{equation}
where, [CMC] is expressed as a mole fraction of the monomers at critical micellar concentration (one needs to use the scaling for CMC ,expressed in mM  using ,CMC=CMC*1E-3; CMC(mole in mole fraction)=CMC/(CMC+55.5), where 55.5 is the molarity of water), the values of $\Delta G $ is between -27.8 to -27.6 $KJ Mole ^{-1}$ (see the break points in Figs. ~\ref{fig:ctab}) or ~\ref{fig:sds}) The values of the free energy change is comparable to what is reported in \cite{gokturk2008effect}. The two intriguing facts that appear is that the enthalpy values derived in \cite{gokturk2008effect} are negative whereas excepting for graphene the values of enthalpy (see Figs. ~\ref{fig:sds} and  ~\ref{fig:ctab}) are positive. It may be noted that since we are doing dilution of micelles into monomer the positive value in ITC experiment actually implies enthalpy of break down. So essentially our results match with the enthalpy signature , implying -ve enthalpy for micelle formation. The case of graphene  however implies that enthalpy change of  micelle formation is actually positive.\\
The conservation of the free energy range varying between -27.8 to -27.6 $KJ Mole ^{-1}$ for a wide class of detergents (triton X data not shown)   suggests that while the free energy remains conserved , there is a enthalpy entropy compensation in the self assembly process  nanosurfaces determining the distribution of the entropic and enthalpic component. The reversal of enthalpy changes in case of graphene (having a positive enthalpy of formation , i.e a negative enthalpy of breakdown) reveals that in presence of graphene the free energy component is dominated by entropic component as with constant $\Delta G =\Delta H -T\Delta S  \le o$ implies that if $\Delta H_{formation}\ge 0 $, $|T\Delta S|_{formation} \ge |\Delta H|_{formation}$ and  $\Delta S_{formation} \ge 0$.  The entropy driven nature of micelle formation is not obligatory in presence of SWNT or MWNTnthalpy of formation is negative. \\
The hydrophobic tails of the surfactant molecules are non-covalently  attached with the hydrophobic surface of SWNTs, MWNTs and graphene. The 
complex forms a highly non-linear structure when the hydrophobic tails are  anchored on SWNTs, MWNTs or graphene surface. A strong repulsive interaction  between the polar head groups of the adhered surfactant molecules on SWNTs,  MWNTs or graphene makes the entire structure highly non-linear. Therefore  the self-assembly is unstable and micelle formation is found to be affected  in presence of such strong hydrophobic materials . The  self-assembly is more unstable with increase concentration of surfactant in  presence of carbon nanomaterials due to strong repulsive interaction between  the polar head groups of the adhered surfactant molecules on SWNTs, MWNTs or  graphene. The molecular self-assembly breaks down with increase in  temperature because the thermal agitation breaks the weak (in terms of energy) self-assembly of different surfactant monomers. 
\section{Conclusion}
The studies imply novel uses of carbon material particularly in areas where bulk thermodynamic properties may  be important. The use of  Carbon Nanotubes as fuel-borne additives in Diesterol blends and their enhanced  performance in combustion and emission as a result has already been pointed out \cite{selvan2014effect}. Similar use of the same in fuel cells have also been pointed out \cite{wang2009electrocatalytic}.  The particular use of graphene as a fuel additive  \cite{sabourin2009functionalized} in say combustion of propellants has also been reported. This paper provides a thermodynamic and hydrodynamic rationale for such usage. Again, according to \cite{zuo2009graphene}, the graphene oxide may facilitate electron transfer of metalloproteins  and the study by\cite{karim2013graphene} show the higher proton conductivity induced by graphene, the use of the same as a bioenegetic processor (e.g. a controllable uncoupler of oxidtative phosphorylation \cite{duch2011minimizing}) seems to be a distinct possibility. The fact that the sign of enthalpy change can be reversed by changing the nanosurface to single wall form to the graphene form is  of particular interest as it shows the intricate relations the geometry and topology have with the thermodynamic behaviour of the self =assembly process.. 


%
%

\begin{acknowledgments}
We wish to acknowledge the support of the author community in using
REV\TeX{}, offering suggestions and encouragement, testing new versions,
\dots.
\end{acknowledgments}

\appendix

\section{Appendixes}

\subsection{Materials}
 
Sodium dodecyl sulphate (SDS), cetyl trimethylammonium bromide (CTAB) and triton X has been purchased from Sigma Aldrich. Single wall carbon nanotubes (SWNTs) and multi wall carbon nanotubes (MWNTs) and graphene have been synthesized by chemical vapor deposition method. Cuvettes for dynamic light scattering study was purchased from Malvern, filters from Millipore and milli Q water from local source. 
\subsection{Synthesis of water solubilized carbon nanomaterials}
About 5mg of the SWNTs, MWNTs and graphene was refluxed in an aqueous solution of neat cold nitric acid (20 ml) for 48-72 hours. A good proportion of the nanomaterials went into the solution. The acidic portion has been removed by serial dilution with milli Q water. Then, the un-dissolved residue was separated by centrifugation and the centrifugate was evaporated to dryness on a water bath to yield a black solid. Further purification of this black solid was performed to yield purified water soluble carbon nanomaterials.
\subsection{Isothermal Titration Calorimetry Study (ITC Study)}
For ITC experiment (GE healthcare ITC 200), aliquots of 40$\mu$L water solubilized SWNTs, MWNTs and graphene (as ligands) have been injected into the calorimeter cell containing 300$\mu$L surfactant micelle.  To avoid an overflow of the initial solution from the cell, the total cell volume has been kept constant. The surfactant dilution have been considered for the calculation of the actual surfactant and functionalized SWNTs, MWNTs or graphene during the experiment. 
\subsection{Photon co-relation study (PCS Study) }
 Photon correlation spectroscopy  have been performed using Malvern 4700 (UK) using a 100-150 mW laser emitting vertically polarized light at a wavelength of 488 nm. The measurements have been carried out at 25$^0$C at first then a temperature dependence study has been also carried out. Before use, the cells have been washed in aqua regia and finally rinsed with Millipore water. The surfactant solutions have been filtered once through a 0.2 micron Millipore filter directly into the cell and sealed until used for removal of dust particles. All measurements (the average scattered intensity and the intensity correlation function) for surfactant micelle and surfactant micelle with carbon nanomaterials have been carried out at least three times for each sample.

\subsection{Matlab Scripts for evaluation of enthalpy of micellar break down}
	function [x,y,cc,H]=enthalpy(data,ninj)\\
	$\%$The volume of the stock solution was 40 $\mu$l  and in every injection 2$\mu$l was added to the stock solution.\\
	$\%$The initial concentration was 10 mM for SDS and 1 mM for CTAB. \\
	$\%$each injection takes 1 sec\\
	close all;\\
	close all;
	cc=[
	10.0000
	9.5000
	9.0250
	8.5738
	8.1451
	7.7378
	7.3509
	6.9834
	6.6342
	6.3025
	5.9874
	5.6880
	5.4036];
	X=data(:,1);\\
	Y=data(:,2);\\
	tau=190;\\
	conc=cc;\\
	conc=conc*1E-3;\\
	plot(data(:,1),data(:,2));
	[x,y]=ginput(ninj);
	H=[];
	H=(y*(1e-6)*4.18)/(0.5*1e-3)*190;
	cc=cc(1:ninj);
	plot(cc,H,'ko')

%

\pagebreak

%

\end{document}